\documentclass[aps,prb,twocolumn,notitlepage,groupedaddress,showpacs]{revtex4-1}

\usepackage{graphicx}

\begin{document}


\title{Static transport properties of random alloys:
       Vertex corrections in conserving approximations}


\author{I. Turek}
\email[]{turek@ipm.cz}
\affiliation{Institute of Physics of Materials,
Academy of Sciences of the Czech Republic,
\v{Z}i\v{z}kova 22, CZ-616 62 Brno, Czech Republic}


\date{\today}

\begin{abstract}
The theoretical formulation and numerical evaluation of the vertex
corrections in multiorbital techniques of theories of electronic
properties of random alloys are analyzed.
It is shown that current approaches to static transport properties
within the so-called conserving approximations lead to the inversion
of a singular matrix as a direct consequence of the Ward identity
relating the vertex corrections to one-particle self-energies.
We propose a simple removal of the singularity for quantities
(operators) with vanishing average values for electron states at
the Fermi energy, such as the velocity or the spin torque;
the proposed scheme is worked out in details in the self-consistent
Born approximation and the coherent potential approximation.
Applications involve calculations of the residual resistivity for
various random alloys, including spin-polarized and relativistic
systems, treated on an \emph{ab initio} level, with particular
attention paid to the role of different symmetries (inversion of
space and time).
\end{abstract}


\maketitle


\section{Introduction\label{s_intr}}

Vertex corrections, encountered in modern Green's function
approaches to interacting electrons \cite{r_2000_gdm} and to
electrons in disordered systems, \cite{r_2000_ag} proved
indispensable in many branches of the solid-state theory and
its applications in materials science.
As an example, let us mention the important role of the vertex
corrections in extensions of the dynamical mean-field theory
for the Hubbard model. \cite{r_2007_tkh}
As concerns transport properties of random alloys, 
the disorder-induced vertex corrections represent the dominating
extrinsic contribution to the anomalous and spin Hall
conductivities of diluted alloys \cite{r_2010_lke, r_2011_lgk}
and they are essential for the residual resistivity of concentrated
binary alloys involving noble and simple metals. \cite{r_1986_sbs} 
Recent \emph{ab initio} studies revealed that the vertex corrections
are significant both for reliable calculations of the Gilbert damping
parameters in disordered magnetic systems \cite{r_2013_mkw,
r_2015_tkd} and for the equivalence of different spin-torque
operators employed in the theory. \cite{r_2015_tkd, r_2015_as_a}
Let us note that the vertex corrections for transport properties
correspond to the scattering-in term in the linearized Boltzmann
equation. \cite{r_1985_whb, r_1999_im}

Basic concepts of the above-mentioned approaches for systems in
equilibrium are one-particle propagators (Green's functions) $G(z)$
and self-energy operators $\Sigma(z)$, where $z$ denotes a complex
energy argument.
The vertex corrections refer to two-particle quantities;
their relation to the one-particle quantities is provided by the
well-known Ward identity. \cite{r_1950_jcw}
This identity is exactly satisfied in exact theories;
for approximate treatments, it represents a check of internal
consistence and it guarantees the conservation of particle number
and energy in the so-called conserving approximations.
General reasons for the validity of the Ward identity can be traced
back to the gauge invariance of the theory both for systems in
equilibrium \cite{r_1957_yt, r_2002_rr} and far from
it. \cite{r_2008_vks}

In the case of noninteracting electrons in random crystalline
alloys, the self-energy $\Sigma(z)$ is related to the configuration
average of the Green's function $\langle G(z) \rangle = \bar{G}(z)$.
The configuration average of a product of two propagators can then
be written as \cite{r_1969_bv}
\begin{equation}
\langle G(z_1) C G(z_2) \rangle = \bar{G}(z_1) C \bar{G}(z_2)
+ \bar{G}(z_1) \Gamma \bar{G}(z_2) ,
\label{eq_gcg}
\end{equation}
where $C$ denotes an arbitrary nonrandom operator (independent of
the particular configuration of the random alloy), the first term on
the r.h.s.\ denotes the coherent contribution and the second term
defines the vertex correction (incoherent part) with the operator
$\Gamma$ depending on $C$ and on both energy arguments,
$\Gamma = \Gamma(z_1, C, z_2)$.
The corresponding Ward identity refers to the special case of unit
operator $C$ ($C=1$) and it has the form
\begin{equation}
\Gamma(z_1, 1, z_2) = - (z_1 - z_2)^{-1} 
 [ \Sigma(z_1) - \Sigma(z_2) ] .
\label{eq_ward}
\end{equation}
The Ward identity is satisfied, e.g., in the self-consistent
Born approximation (SCBA) \cite{r_1958_sfe, r_2008_vks}
and in the coherent potential approximation
(CPA); \cite{r_1967_ps, r_1969_bv, r_1974_ekl}
the former is suitable for weak static fluctuations of the random
one-particle Hamiltonian while the latter can be applied even to
strong fluctuations but with uncorrelated contributions of different
lattice sites.

The dependence of the vertex correction $\Gamma(z_1, C, z_2)$ on
the operator $C$ is linear and finding the $\Gamma$ for a given $C$
is equivalent to solving a Bethe-Salpeter equation. \cite{r_1969_bv}
Corresponding numerical procedures have been developed for systems
featured by a finite number of orbitals per lattice site and they
have also been worked out in \emph{ab initio} techniques, such as
the Korringa-Kohn-Rostoker (KKR) method \cite{r_2000_ag, r_1985_whb}
or the tight-binding linear muffin-tin orbital (TB-LMTO) method.
\cite{r_2006_ctk} 
For zero-temperature static transport properties, the energy
arguments $z_1$ and $z_2$ in Eq.~(\ref{eq_gcg}) acquire values
$E_\mathrm{F} \pm i0$, where $E_\mathrm{F}$ denotes the alloy Fermi
energy.
For $z_1 = E_\mathrm{F} + i0$ (retarded propagator and self-energy)
and $z_2 = E_\mathrm{F} - i0$ (advanced quantities),
the denominator in Eq.~(\ref{eq_ward}) approaches zero, whereas the
difference of the self-energies remains finite as long as the Fermi
energy lies inside the spectrum, i.e., for metallic alloys.
The divergence of the r.h.s.\ of Eq.~(\ref{eq_ward}) in this case
proves that the linear relation between $C$ and $\Gamma$ is singular.

The singular behavior of the vertex corrections for small energy
and momentum transfers has been discussed by a number of authors for
systems with electron interactions \cite{r_2002_rr, r_1989_tt} as well
as for noninteracting electrons in disordered alloys especially in the
context of Anderson localization. \cite{r_1980_vw, r_2009_vj}
Existing first-principles calculations of transport properties of
random alloys often employ a finite imaginary part added to both
energy arguments, $z_{1,2} = E_\mathrm{F} \pm i \eta$, where $\eta$
is a small positive quantity \cite{r_2015_tkd, r_2015_as_a}
which can be interpreted as an additional broadening of electron
energy levels due to unspecified mechanisms ignored in the theory
(structural defects, phonons). \cite{r_2009_ktt, r_2015_kdt_a}
From the numerical point of view, the use of a finite $\eta$ removes
the singularity in the vertex corrections. 
However, with a recent progress in the realistic inclusion of
temperature-induced phonons and magnons on the transport
properties, \cite{r_2015_emc}
the introduction of any artificial broadening mechanism does not
seem desirable and the problem of reliable calculations for $\eta=0$
should thus be solved in a different way.
It is the purpose of this paper to propose a practical scheme in
this direction and to show its efficiency in calculations of the
residual resistivity of random metallic alloys.
Since the removal of the general singularity due to the Ward identity
(\ref{eq_ward}) can be simplified (or complicated) by the symmetries
of the considered system, such as its invariance with respect to
space and time inversion, their relevance will also be discussed in
the text.

\section{Theoretical formalism\label{s_form}}

In the following, we consider random alloys on a nonrandom
crystal lattice with sites labelled by an index $\mathbf{R}$.
The effective one-electron Hamiltonian $H$ is represented in an
orthonormal orbital basis $| \mathbf{R}L \rangle$ by a matrix
$H_{\mathbf{R}_1 L_1, \mathbf{R}_2 L_2}$, where $L$, $L_1$ and $L_2$
label the atomic-like orbitals. 
The random Hamiltonian can be written as $H = H_0 + D$, where
$H_0$ denotes the nonrandom part, while the random part $D$ can be
written as a lattice sum of individual site-contributions,
$D = \sum_\mathbf{R} D_\mathbf{R}$. 
We assume that each term $D_\mathbf{R}$ depends only on the atomic
species occupying the site $\mathbf{R}$ and that its average value
vanishes, $\langle D_\mathbf{R} \rangle = 0$, and we neglect any
correlations of occupations of different lattice sites.
Moreover, we assume that each contribution $D_\mathbf{R}$ is
localized to its own site:
$(D_\mathbf{R})_{\mathbf{R}_1 L_1, \mathbf{R}_2 L_2} = 
\delta_{\mathbf{R}_1 \mathbf{R}} \delta_{\mathbf{R}_2 \mathbf{R}}
D_{\mathbf{R}, L_1 L_2}$. 
The configuration average of the Green's function 
$G(z) = (z - H)^{-1}$ can be written in terms of the self-energy
$\Sigma(z)$ as $\bar{G}(z) = (z - H_0 - \Sigma(z) )^{-1}$.

In the SCBA, \cite{r_1958_sfe} the self-energy is defined by the
condition $\Sigma(z) = \langle D \bar{G}(z) D \rangle$. 
Under the above assumptions, the total self-energy $\Sigma(z)$
reduces to a lattice sum $\Sigma(z) = \sum_\mathbf{R}
 \Sigma_\mathbf{R}(z)$, where the site-contributions
$\Sigma_\mathbf{R}(z)$ are localized, given explicitly by
$\Sigma_\mathbf{R}(z) = \langle D_\mathbf{R} \bar{G}(z)
D_\mathbf{R} \rangle$. 
The SCBA-vertex correction $\Gamma(z_1, C, z_2)$ in
Eq.~(\ref{eq_gcg}) can be found from the 
condition \cite{r_2008_vks, r_1958_sfe}
\begin{equation}
\Gamma = \left\langle D \bar{G}(z_1) ( C + \Gamma )
 \bar{G}(z_2) D \right\rangle , 
\label{eq_vcscba}
\end{equation}
which implies that the complete $\Gamma$ reduces again to
a lattice sum, $\Gamma = \sum_\mathbf{R} \Gamma_\mathbf{R}$,
of localized site-contributions $\Gamma_\mathbf{R}$.
In order to convert Eq.~(\ref{eq_vcscba}) into an explicit set
of linear equations for the quantities $\Gamma_\mathbf{R}$ in
multiorbital techniques, one can introduce composed orbital
indices $\Lambda = (L, L')$, $\Lambda_1 = (L_1, L_1')$, etc.\
together with vector components
$\Gamma_{\mathbf{R}\Lambda} = \Gamma_{\mathbf{R}, L L'}$ and
$\zeta_{\mathbf{R}\Lambda} = [ \bar{G}(z_1) C \bar{G}(z_2) 
]_{\mathbf{R} L, \mathbf{R} L'}$ and with matrix elements
\begin{eqnarray}
\psi_{\mathbf{R}_1 \Lambda_1 , \mathbf{R}_2 \Lambda_2} & = &
\bar{G}_{\mathbf{R}_1 L_1 , \mathbf{R}_2 L_2} (z_1)
\bar{G}_{\mathbf{R}_2 L_2' , \mathbf{R}_1 L_1'} (z_2) ,
\nonumber\\
\mathcal{L}_{\mathbf{R}_1 \Lambda_1 , \mathbf{R}_2 \Lambda_2}
 & = & 
\delta_{\mathbf{R}_1 \mathbf{R}_2}
\left\langle D_{\mathbf{R}_1, L_1 L_2}
 D_{\mathbf{R}_1, L_2' L_1'} \right\rangle . 
\label{eq_deflp}
\end{eqnarray}
The condition (\ref{eq_vcscba}) can then be written in an obvious
matrix notation as
$\Gamma = \mathcal{L} ( \zeta + \psi \Gamma )$, or
\begin{equation}
\Delta \Gamma = \zeta, \qquad \Delta = \mathcal{L}^{-1} - \psi .
\label{eq_defdel}
\end{equation}
If the matrix $\Delta_{\mathbf{R}_1 \Lambda_1 , \mathbf{R}_2
 \Lambda_2}$ is nonsingular, the vertex corrections
$\Gamma_{\mathbf{R}\Lambda}$ can easily be obtained.
The techniques for solving Eq.~(\ref{eq_defdel}) in the case of
translationally invariant operators $C$ and extended systems 
can be found elsewhere. \cite{r_1985_whb, r_2006_ctk}

Let us consider the matrix $\Delta$ (\ref{eq_defdel}) for 
$z_1 = E_\mathrm{F} + i0$ and $z_2 = E_\mathrm{F} - i0$,
and let us denote by $\tilde{\Delta}$ the same matrix for
$z_1 = E_\mathrm{F} - i0$ and $z_2 = E_\mathrm{F} + i0$.
As mentioned in Section \ref{s_intr}, these matrices are
singular: as a consequence of the Ward identity (\ref{eq_ward}),
it holds $\Delta N = 0$ and $\tilde{\Delta} N = 0$, where the
nonzero vector $N = \{ N_{\mathbf{R}\Lambda} \}$ has components 
\begin{equation}
N_{\mathbf{R}\Lambda} = 
\Sigma_{\mathbf{R}, L L'}(E_\mathrm{F} + i0) - 
\Sigma_{\mathbf{R}, L L'}(E_\mathrm{F} - i0) . 
\label{eq_defn}
\end{equation}
If we introduce $\tilde{\Lambda} = ( L' , L )$ for
$\Lambda = (L, L')$, then one can prove easily
$\tilde{\Delta}_{\mathbf{R}_1 \Lambda_1 , \mathbf{R}_2 
\Lambda_2} = \Delta_{\mathbf{R}_2 \tilde{\Lambda}_2 , 
\mathbf{R}_1 \tilde{\Lambda}_1}$, and the condition
$\tilde{\Delta} N = 0$ can be rewritten as
\begin{equation}
\sum_{\mathbf{R}_1 \Lambda_1} N_{\mathbf{R}_1 \tilde{\Lambda}_1}
\Delta_{\mathbf{R}_1 \Lambda_1 , \mathbf{R}_2 \Lambda_2} = 0 .
\label{eq_nd}
\end{equation}
This relation yields immediately a necessary condition for the
existence of the solution of Eq.~(\ref{eq_defdel}): 
\begin{equation}
\sum_{\mathbf{R} \Lambda} N_{\mathbf{R} \tilde{\Lambda} }
 \zeta_{\mathbf{R} \Lambda} = 0 .
\label{eq_nz}
\end{equation}
The last rule can be reformulated as follows.
If we abbreviate $\Sigma^\pm = \Sigma(E_\mathrm{F} \pm i0)$
and $\bar{G}^\pm = \bar{G}(E_\mathrm{F} \pm i0)$ and denote the
trace by $\mathrm{Tr}$, then Eq.~(\ref{eq_nz}) is equivalent to
\begin{eqnarray}
0 & = & \mathrm{Tr} \{ (\Sigma^+ - \Sigma^-) \bar{G}^+ C \bar{G}^- \}
\nonumber\\
  & = & \mathrm{Tr} \{ \bar{G}^- (\Sigma^+ - \Sigma^-) \bar{G}^+ C \}
\nonumber\\
  & = & \mathrm{Tr} \{ ( \bar{G}^+ - \bar{G}^- ) C \} ,
\label{eq_avc}
\end{eqnarray}
where in the last step the Dyson equation relating mutually both
propagators $\bar{G}^\pm = (E_\mathrm{F} - H_0 - \Sigma^\pm)^{-1}$
has been used.
The obtained condition (\ref{eq_avc}) has a transparent physical
interpretation: it means that the average value of the operator $C$
for electron states at the Fermi energy vanishes.
The condition (\ref{eq_nz}) for the existence of the solution of
Eq.~(\ref{eq_defdel}) is thus satisfied by usual velocity operators
entering the Kubo formula for the conductivity tensor.
Another operator $C$ satisfying this condition is the spin-torque
operator in ferromagnets with the magnetization vector in an
equilibrium direction, i.e., pointing along the easy or hard axis.
It should be noted that $N_{\mathbf{R} \tilde{\Lambda} } =
 - N^*_{\mathbf{R} \Lambda }$ which means that the condition
(\ref{eq_nz}) represents an orthogonality relation between the
vectors $\zeta = \{ \zeta_{\mathbf{R} \Lambda} \}$ and 
$N = \{ N_{\mathbf{R} \Lambda} \}$.
The solution of Eq.~(\ref{eq_defdel}) for the vertex corrections
$\Gamma = \{ \Gamma_{\mathbf{R} \Lambda} \}$ can be now performed
in the vector space orthogonal to the vector
$N = \{ N_{\mathbf{R} \Lambda} \}$ (\ref{eq_defn}), which removes
the effect of singularity of the matrix $\Delta$ due to the relation
$\Delta N = 0$.
This solution can be written formally as
\begin{equation}
\Gamma = ( \Pi / \Delta ) \zeta ,
\label{eq_gamma}
\end{equation}
where $\Pi$ denotes the projection operator on the vector space
orthogonal to the vector $N$ and where the L\"owdin's symbol
$( \Pi / \Delta )$ for the restricted inverse has been
used. \cite{r_1962_pol}
This restriction of the vector space for the vertex corrections
is an analogy to the restriction due to conservation of the number
of particles encountered in exact solutions of integral equations
of the linearized Boltzmann theory. \cite{r_2009_vks_a}
Let us note for completeness that the solution of
Eq.~(\ref{eq_defdel}) for the unknown vector $\Gamma$ is not unique
(in the considered case of $z_1 = E_\mathrm{F} + i0$ and
$z_2 = E_\mathrm{F} - i0$), but it is defined up to a term parallel
to the vector $N$.
This ambiguity can be removed by evaluating the limit of
$\Gamma(E_\mathrm{F} + i\eta, C, E_\mathrm{F} - i\eta)$
for $\eta \to 0$. 
However, the additional contribution to $\Gamma$ (parallel to $N$)
has no effect on values of typical linear-response coefficients
$\mathrm{Tr} \langle G(z_1) C G(z_2) C' \rangle$, where 
$z_1 = E_\mathrm{F} + i0$ and $z_2 = E_\mathrm{F} - i0$ and where
both nonrandom operators $C$ and $C'$ satisfy the
condition (\ref{eq_avc}).

The above approach removes the divergence of the vertex corrections
due to the Ward identity and the conservation of the number of
particles of the whole system.
However, particular systems and models can have special properties
which call for more sophisticated treatments, or offer simpler
solutions of the problem.
A detailed analysis of these special cases goes beyond the scope
of this work; let us mention only two examples here. 
First, let us consider the case of a random ferromagnetic alloy in
models without spin-orbit interaction.
The two spin channels are decoupled from each other and,
consequently, there exist two linearly independent vectors $N$
(\ref{eq_defn}), $N^\uparrow$ and $N^\downarrow$, satisfying the
relation $\Delta N = 0$.
The removal of the singularity of $\Delta$ leads naturally to a
subspace orthogonal to both vectors $N^\uparrow$ and $N^\downarrow$,
whereas a simpler solution would be a separate treatment of both
spin channels in the spirit of the two-channel model of electron
transport. \cite{r_1964_nfm}
Second, let us consider the conductivity tensor of random systems
invariant to space inversion, such as homogeneous solid solutions
on bcc or fcc lattices.
Since the unperturbed Hamiltonian $H_0$, the random perturbations
$D_\mathbf{R}$, the average Green's functions $\bar{G}(z)$ and the
self-energies $\Sigma_\mathbf{R}(z)$ are even quantities with respect
to space inversion, whereas the velocity operator $C$ and the
corresponding vertex corrections $\Gamma_\mathbf{R}$ are odd,
an elementary group theory \cite{r_1960_vh} can be applied to
Eq.~(\ref{eq_defdel}).
The singular behavior due to the Ward identity (\ref{eq_ward}) is
then confined to the even subspace that is decoupled from the odd
subspace, which leads automatically to nonsingular vertex
corrections to the conductivity tensor.

\begin{figure}
\includegraphics[width=0.40\textwidth]{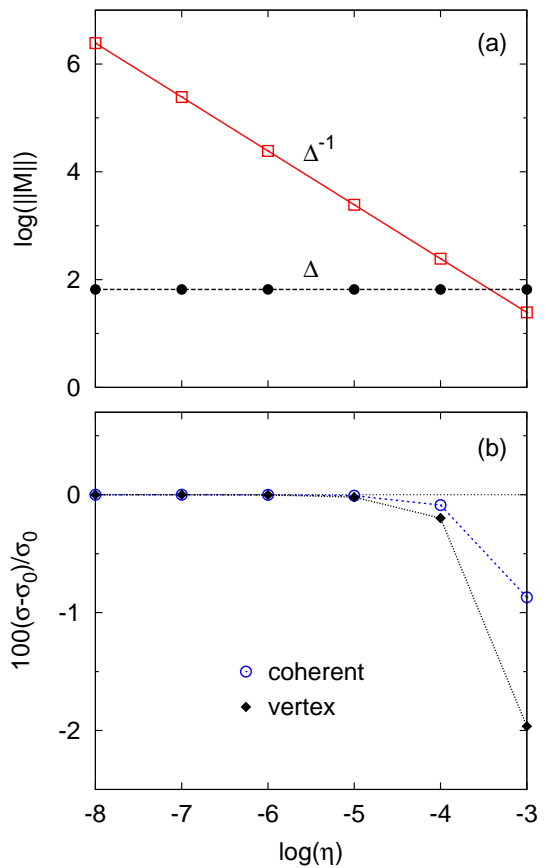}
\caption{(Color online)
Quantities related to the one-dimensional tight-binding model in
the SCBA as functions of imaginary part $\eta$ of the energy: 
(a) the norm of matrices $\Delta$ and $\Delta^{-1}$,
(b) the relative deviations of the coherent and incoherent (vertex)
contributions to the residual conductivity $\sigma$ with respect to
their values for $\eta=0$.
\label{f_tb1d}}
\end{figure}

Let us illustrate the developed formalism by a simple example,
namely, by the application to a hypothetical one-dimensional
tight-binding model of a random alloy treated in the SCBA.
A similar model was studied by Butler using the KKR-CPA
theory, \cite{r_1985_whb} which however was limited to the case
of symmetric potentials of both atomic species, i.e., to the
case with space-inversion symmetry mentioned above.
Here we consider a model with two atomic-like orbitals per site,
featured by a symmetric ($s$ orbital) and an antisymmetric 
($p$ orbital) shapes.
The lattice site occupy a one-dimensional Bravais lattice with a
lattice parameter $a=1$; the unperturbed Hamiltonian $H_0$ and the
nonrandom velocity operator are defined in terms of on-site
atomic levels ($\epsilon_s = -0.1$, $\epsilon_p = -0.2$, both
values given with respect to the Fermi energy) and the
nearest-neighbor hopping integrals ($W_{ss} = 0.6$,
$W_{sp} = -0.25$, $W_{pp} = 0.4$).
The matrix elements of the random on-site perturbations have been
chosen to describe nonsymmetric potentials ($D_{ss} = \pm 0.15$,
$D_{sp} = \pm 0.3$, $D_{pp} = \pm 0.2$), where the two signs
refer to two atomic species with equal concentrations.
The evaluation of the residual conductivity using the Kubo-Greenwood
formula \cite{r_2000_ag, r_1957_rk, r_1958_dag} has been carried out
with complex energies $z_{1,2} = E_\mathrm{F} \pm i\eta$ ($\eta > 0$)
without any modification in solving the vertex corrections according
to Eq.~(\ref{eq_defdel}) as well as with real energy arguments
$z_{1,2} = E_\mathrm{F} \pm i0$ according to the developed general
regularization procedure (\ref{eq_gamma}). 
The results are shown in Fig.~\ref{f_tb1d}.
This simple case leads to a $4 \times 4$ matrix $\Delta$;
its Frobenius (Hilbert-Schmidt) matrix norm $|| \Delta ||$ together
with $|| \Delta^{-1} ||$ are displayed in Fig.~\ref{f_tb1d}(a) as
functions of $\eta$.
The diverging trend of $|| \Delta^{-1} ||$ for $\eta \to 0$ proves
the singularity mentioned above.
The calculation of the incoherent (vertex) part of the conductivity
for $\eta=0$ with help of Eq.~(\ref{eq_gamma}) involves inversion
of a $3 \times 3$ matrix.
Its matrix norm coincides with that of the original $4 \times 4$
matrix $\Delta$, but the norm of its inverse is finite, 
$|| \Pi / \Delta || \approx 7 \times 10^{-2}$ in the
present case, which is much smaller than the big values of
$|| \Delta^{-1} ||$ for the positive values of $\eta$ shown in
Fig.~\ref{f_tb1d}.
The regularization procedure based on Eq.~(\ref{eq_gamma}) thus
allows one not only to obtain directly the conductivity for $\eta=0$,
but also to improve substantially the numerical stability of the
original linear problem (\ref{eq_defdel}).
The relation of the coherent ($\sigma^\mathrm{coh}$) and vertex
($\sigma^\mathrm{vc}$) parts of the conductivity for nonzero $\eta$
to their limiting values for $\eta=0$ ($\sigma^\mathrm{coh}_0 = 38.8$,
$\sigma^\mathrm{vc}_0 = 0.87$) is depicted in Fig.~\ref{f_tb1d}(b);
it documents a quick convergence of both contributions.

The presented removal of the singularity is not confined to the SCBA;
its generalization to the CPA is straightforward, since the linear
condition (\ref{eq_defdel}) for the vertex corrections has the same
form with a slightly modified matrix $\Delta$. \cite{r_2006_ctk}
Let us mention for completeness that the underlying idea is
independent on the specific approximation used as well as on details
of the potential fluctuations, so that even delocalized perturbations
$D_\mathbf{R}$ with arbitrary correlations among different lattice
sites are allowed.
This follows from the identity
\begin{equation}
\mathrm{Tr} \langle G(z_1) C G(z_2) \rangle =
\mathrm{Tr} \langle G(z_2) G(z_1) C \rangle ,
\label{eq_ide}
\end{equation}
valid for any nonrandom $C$ and arbitrary arguments $z_{1,2}$ owing
to the cyclic property of trace.
By writing the l.h.s.\ in terms of the vertex corrections
$\Gamma$ (\ref{eq_gcg}) and using the Ward identity (\ref{eq_ward})
on the r.h.s., one obtains easily a relation
\begin{eqnarray}
 & & (z_2 - z_1) 
 \mathrm{Tr} \{ \bar{G}(z_1) \Gamma \bar{G}(z_2) \} 
\nonumber\\
 & & = 
 \mathrm{Tr} \{ \bar{G}(z_2) [ \Sigma(z_1) - \Sigma(z_2) ]
\bar{G}(z_1) C \} .
\label{eq_auxr}
\end{eqnarray}
The requirement of a nonsingular $\Gamma$ in the limit
$z_1 \to E_\mathrm{F} + i0$ and $z_2 \to E_\mathrm{F} - i0$
yields immediately the condition (\ref{eq_avc}) for the vanishing 
average of $C$ at the Fermi energy.

Let us conclude this section by several remarks.
First, the above discussed singularity is always present in the
matrix $\Delta$ (for $z_1 = E_\mathrm{F} + i0$,
$z_2 = E_\mathrm{F} - i0$) which prevents its direct inverse.
This matrix depends only on the Hamiltonian $H$ of the random alloy.
This singularity, however, is suppressed in the incoherent part
of a particular transport coefficient
$\mathrm{Tr} \langle G^+ C G^- C' \rangle$,
where $G^\pm = G(E_\mathrm{F} \pm i0)$, if both nonrandom operators
$C$ and $C'$ satisfy the condition (\ref{eq_avc}).
The developed scheme based on Eq.~(\ref{eq_gamma}) enables one to
avoid the singularity of $\Delta$ in obtaining the incoherent part
of the transport coefficient.
Second, the applicability of the presented formalism is not confined
to zero-temperature properties where the Fermi energy plays the
central role, but it can easily be extended to finite temperatures. 
In the latter case, the Fermi energy $E_\mathrm{F}$ has to be
replaced by a real energy variable and the resulting transport
coefficients (e.g., conductivity or Seebeck coefficient) are
obtained by the corresponding energy integration according to the
Mott formula. 
Third, the singularity of the matrix $\Delta$ is in general
encountered only for the complex arguments $z_1$ and $z_2$
approaching the same real energy (inside the alloy spectrum)
from opposite sides.
In particular, the treatment of the so-called Fermi-sea
term \cite{r_2014_tkd, r_2015_kce} appearing in the Bastin
formula, \cite{r_1971_blb} where both complex arguments lie
simultaneously in the upper or lower halfplane, does
not lead to the discussed singularity.
Similarly, the case of various frequency-dependent quantities
(dynamical susceptibilities, optical conductivities) for a finite
frequency $\omega$, where both energy arguments are separated by
$\hbar \omega$, \cite{r_2000_ag} does not require any special care
in evaluation of the vertex corrections.

\section{Applications to realistic models\label{s_appl}}

Let us turn finally to applications of the developed procedure
in \emph{ab initio} studies of transport properties of random
metallic alloys performed in the CPA.
In the following, we will discuss the calculation of the
residual resistivity as a basic transport property for fcc
Ag$_{0.5}$Pd$_{0.5}$ and bcc Fe$_{0.8}$Al$_{0.2}$ solid
solutions and for a diluted magnetic semiconductor, namely, GaAs
doped by 8\% Mn atoms substituting Ga atoms.
This limited choice of systems includes both nonmagnetic (Ag-Pd)
and ferromagnetic (Fe-Al, Mn-doped GaAs) alloys as well as
systems with (Ag-Pd, Fe-Al) and without (Mn-doped GaAs) space
inversion symmetry.
Moreover, we applied both scalar-relativistic \cite{r_2002_tkd,
r_2004_tkd} and fully relativistic \cite{r_2012_tkd} versions of
the transport theory in the TB-LMTO method; in all cases the
valence basis comprised $s$-, $p$- and $d$-like orbitals.
The site-diagonal self-energy $\Sigma_{\mathbf{R}, L L'}(z)$ 
has been replaced by the coherent potential functions
$\mathcal{P}_{\mathbf{R}, L L'}(z)$ and other quantities of
Section \ref{s_form} by their LMTO counterparts according to
Appendix of Ref.~\onlinecite{r_2006_ctk}.
The very small Fermi-sea contribution to the conductivity
tensor \cite{r_2014_tkd} has been omitted here.
Note that the presence of spin-orbit interaction allows one
to distinguish systems with (Ag-Pd) and without (Fe-Al, Mn-doped
GaAs) time-inversion symmetry.

\begin{figure}
\includegraphics[width=0.40\textwidth]{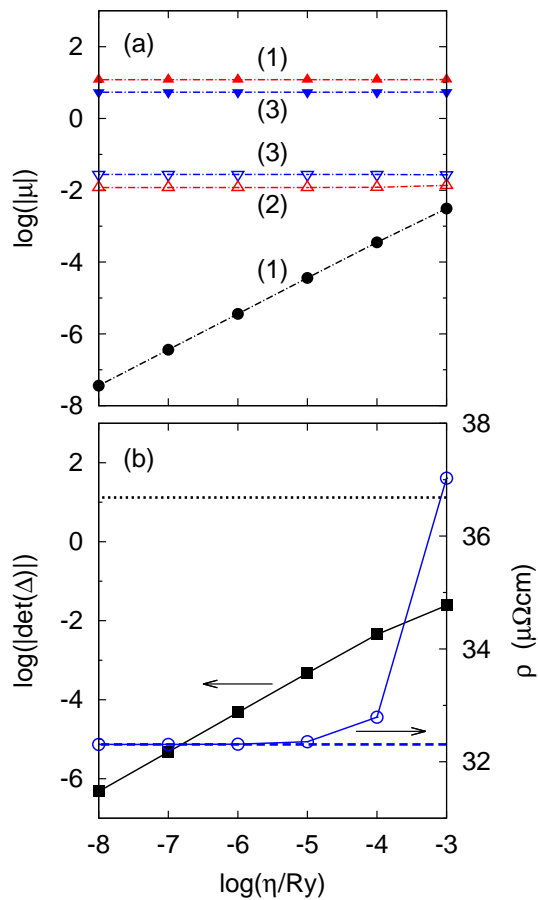}
\caption{(Color online)
Analysis of the case of fcc Ag$_{0.5}$Pd$_{0.5}$ alloy in the
scalar-relativistic approximation:
(a) Absolute values of selected eigenvalues $\mu_i$ of the matrix
$M$ as functions of the imaginary part $\eta$ of energy, see
text for details.
The degeneracies of the eigenvalues are given in parenthesis.
(b) Absolute value of determinant of the matrix $\Delta$ (left
scale, full squares) and of the residual resistivity $\rho$ (right
scale, open circles) as functions of $\eta$.
The dotted vertical line marks absolute value of determinant of
the restricted matrix $\Delta$ and the dashed vertical line
denotes the value of $\rho$ for $\eta=0$.
\label{f_sragpd}}
\end{figure}

The most detailed analysis has been performed for the
scalar-relativistic calculation of the Ag$_{0.5}$Pd$_{0.5}$ alloy. 
Since the norm of matrices $\Delta$ and $\Delta^{-1}$ represents
incomplete information about the stability of the set of linear
equations (\ref{eq_defdel}), we have studied also the determinant
of the matrix $\Delta$ and its eigenvalues.
The matrix $\Delta$ (for $z_1 = E_\mathrm{F} + i\eta$ and
$z_2 = E_\mathrm{F} - i\eta$) is not Hermitean; however, for a system
without spin polarization and spin-orbit interaction (and with the
orbital index $L$ labelling real spherical harmonics), the matrix
$M$ with elements $M_{\mathbf{R}_1 \Lambda_1 , \mathbf{R}_2 
\Lambda_2} = \Delta_{\mathbf{R}_1 \tilde{\Lambda}_1 , 
\mathbf{R}_2 \Lambda_2}$ is Hermitean, so that its all eigenvalues
$\mu_i$ are real and they can be obtained by standard means.
(In fact, only the lattice Fourier transform of both matrices $M$
and $\Delta$ for zero reciprocal-space vector has to be considered,
see Ref.~\onlinecite{r_2006_ctk}.)
Note that the matrices $M$ and $\Delta$ differ only by a permutation
of their rows, hence the numerical stability of the system
(\ref{eq_defdel}) can be assessed equally well by inspecting any of
them.
Selected eigenvalues $\mu_i$ of the matrix $M$ as functions of the
imaginary part $\eta$ of energy arguments $z_{1,2}$ are displayed in
Fig.~\ref{f_sragpd}(a).
The spectrum of $M$ contains a nondegenerate eigenvalue with the
magnitude roughly proportional $\eta$ (marked by full circles).
The other eigenvalues are essentially independent of $\eta$; only
the lowest/highest negative (full/open triangles down) and
the lowest/highest positive (open/full triangles up) eigenvalues
are shown in Fig.~\ref{f_sragpd}(a).
The degeneracies of all eigenvalues equal 1, 2, or 3, in agreement
with dimensions of irreducible representations of the full cubic
point group. \cite{r_1960_vh, r_1957_gfk}
The nondegenerate eigenvalue approaching zero for $\eta \to 0$
(full circles) proves the existence of a single linearly independent
vector $N$ satisfying $\Delta N = 0$ for $\eta=0$, so that the
restricted inversion in Eq.~(\ref{eq_gamma}) can be performed.

As a consequence of the above trends of the eigenvalues $\mu_i$,
the absolute value of the determinant of matrix $\Delta$ is
proportional $\eta$, as shown in Fig.~\ref{f_sragpd}(b), and
it vanishes for $\eta=0$.
The values of the residual resistivity $\rho$ for finite values
of $\eta$ converge rapidly to the limiting value obtained for
$\eta=0$ with help of Eq.~(\ref{eq_gamma}).
Moreover, the absolute magnitude of the determinant of the
restricted matrix $\Delta$ is several orders of magnitude larger
than that of the original matrices $\Delta$, see
Fig.~\ref{f_sragpd}(b), which indicates improved numerical
stability in analogy to the model case (Section~\ref{s_form}).
Qualitatively identical results have also been obtained for the
conducting majority-spin channel of Mn-doped GaAs in the absence of
spin-orbit interaction as a system without space-inversion symmetry
(not shown here).

\begin{figure}
\includegraphics[width=0.40\textwidth]{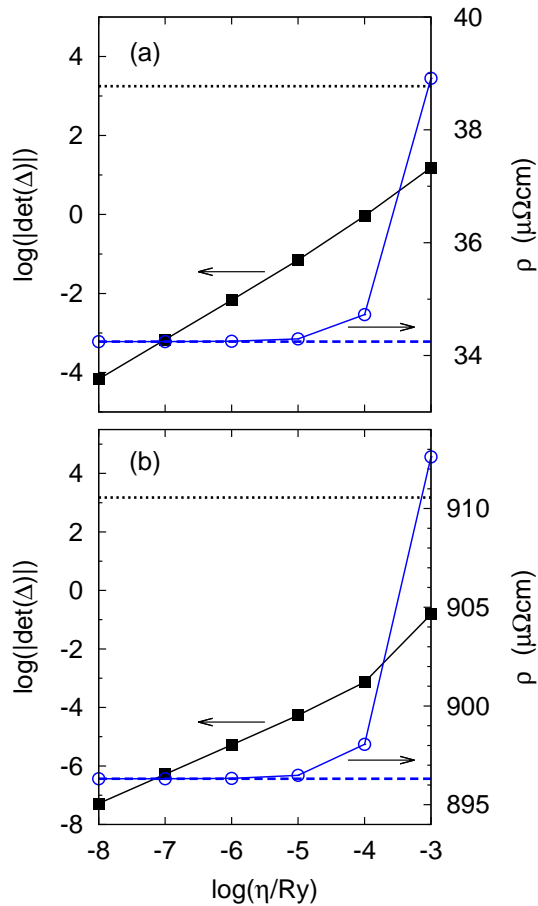}
\caption{(Color online)
The same as in Fig.~\ref{f_sragpd}(b) for the fully relativistic
treatment of fcc Ag$_{0.5}$Pd$_{0.5}$ (a) and
(Ga$_{0.92}$Mn$_{0.08}$)As (b).
\label{f_frx}}
\end{figure}

Results of calculations for systems with spin-orbit interaction are
summarized in Fig.~\ref{f_frx}.
The nonmagnetic random fcc Ag$_{0.5}$Pd$_{0.5}$ alloy
[Fig.~\ref{f_frx}(a)] represents a case with full cubic and
time-inversion symmetry.
All one-electron eigenvalues of pure crystals of such systems have
even degeneracies; \cite{r_1960_vh, r_1957_gfk} the order of
singularity of the matrix $\Delta$ for $\eta \to 0$ requires thus
special attention.
The data displayed in Fig.~\ref{f_frx}(a) prove a proportionality
between $|\mathrm{det} (\Delta)|$ and $\eta$, which means that the
restricted inverse in Eq.~(\ref{eq_gamma}) is nonsingular and it
can be performed similarly with the previous spinless case.
The convergence of the residual resistivity $\rho$ for $\eta \to 0$
and the improvement of numerical stability due to the restricted
inverse are also independent on spin-orbit interaction, see
Fig.~\ref{f_sragpd}(b) and Fig.~\ref{f_frx}(a).

The ferromagnetic Mn-doped GaAs with magnetization pointing along
$z$ axis [Fig.~\ref{f_frx}(b)] represents an opposite case, namely,
a system without the time-inversion symmetry and with the point
group reduced to $S_4$. 
The proportionality between $|\mathrm{det} (\Delta)|$ and $\eta$ can
again be seen in Fig.~\ref{f_frx}(b), which proves the applicability
of Eq.~(\ref{eq_gamma}) also in this case, as confirmed by the
calculated resistivities $\rho$ and their convergence.
Let us mention that a qualitatively identical behavior has been
obtained for the random ferromagnetic bcc Fe$_{0.8}$Al$_{0.2}$ alloy
with spin-orbit interaction and with magnetization pointing along
$[100]$, $[110]$ and $[111]$ directions (not shown here).

The results of calculations for the selected systems allow one to
conclude that the simple restricted inverse (\ref{eq_gamma}) is
generally applicable for realistic models of random systems
irrespective of their geometrical and time-inversion symmetries;
the only exceptions seem to be cases with very special symmetries,
such as, e.g., ferromagnets with omitted spin-orbit interaction
(see Section \ref{s_form}).

\section{Conclusions\label{s_conc}}

This study addressed the problem of removing a singularity in the
vertex corrections that is encountered in the case of zero energy
and momentum transfer, which is relevant for the static response of
random alloys to homogeneous external perturbations.
The singularity reflects basic conservation laws as expressed by the
Ward identity satisfied by standard conserving approximations (SCBA,
CPA).
This identity also provides a key for a simple solution of the
problem for transport properties, which involve operators (velocity,
spin torque) with zero average values for electron states at the
Fermi energy. 
The developed formalism, worked out in multiorbital techniques
applicable to realistic models of random alloys, is based on a
restriction of the vector space for the vertex corrections; the
dimension of the original vector space has to be reduced by unity,
which leads as a rule to a regular matrix inversion.
In principle, one cannot exclude more complex situations, which
require more sophisticated solutions, especially for systems
possessing very special symmetries.
A complete solution to this problem (if it exists at all) goes
beyond the scope of this work; however, usual symmetry operations
of most alloy systems, such as inversion of time and space as well
as rotations and reflections, do not call for any modification of
the suggested approach.
The illustrating examples in this work have been confined to 
electrical resistivity, but extensions to other transport
quantities, such as, e.g., the Gilbert damping 
parameters \cite{r_2015_tkd} or spin-orbit torques induced by
external electric fields, \cite{r_2014_fbm} can be done in a
straightforward manner.

\begin{acknowledgments}
The author acknowledges financial support from the Czech Science
Foundation (Grant No.\ 15-13436S).
\end{acknowledgments}



\providecommand{\noopsort}[1]{}\providecommand{\singleletter}[1]{#1}%

\end{document}